\documentstyle[12pt]{article}

\begin{document}

\begin{titlepage}

\title{Distribution of Heights in the Abelian Sandpile Model
on the Husimi Lattice.}

\author{Vl.V.~Papoyan\thanks{E-mail: papoyanv@theor.jinrc.dubna.su} \\[2mm]
and \\[2mm]
R.R.~Shcherbakov\thanks{E-mail: shcher@thsun1.jinr.dubna.su}
\thanks{Permanent address: Theoretical Department, 
Yerevan Physics Institute, 
Alikhanian Br., Str. 2, Yerevan, 375036, Armenia.}\\[2mm]
{\small \sl Bogoliubov Laboratory of Theoretical Physics,} \\
{\small \sl JINR, 141980 Dubna, Russia.}
}

\date{}

\maketitle

\begin{abstract}
An Abelian sandpile model is considered on the Husimi lattice
of triangles with an arbitrary coordination number $q$.
Exact expressions for the distribution of height probabilities
in the Self-Organized Critical state are derived.
\end{abstract}

\thispagestyle{empty}
\end{titlepage}

\section{Introduction.}

To explain the temporal and spatial scaling in dynamical
dissipative systems being in the {\sl Self-\-Organized Critical}
(SOC) state, Bak, Tang, and Weisenfeld have introduced cellular
automaton models known as  {\sl sandpiles}~\cite{BTW}.

Latter on, Dhar has studied sandpile models which can be
described by the Abelian group~\cite{Dhar}.
For these {\sl Abelian sandpile models} (ASM's) some
ensemble-\-average quantities such as the total number of
configurations in the SOC state, distribution of height
probabilities, two-point correlation functions, etc., have
been calculated exactly on the square lattice~[3-6].

Unfortunately, for the sandpile models formulated on the
standard plane lattices it seems to be rather difficult
to formulate a direct theoretical approach leading to the
solution in closed form.

An appreciable insight can be obtained, however, by examining
pseudolattices such as the {\sl Bethe lattice, Husimi lattice},
and their generalizations.

Using this approach, Dhar and Majumdar~\cite{DM} have obtained
exactly a more complete set of sandpile characteristics for the
Bethe lattice.

The aim of this article is to analyze the distribution of height
probabilities of the ASM on the Husimi lattice with an arbitrary
coordination number $q$. We show that this model can be solved
exactly by using the fractal structure of the lattice.

The outline of the article is as follows:
in the next section, we define the  lattice and the ASM on it.
In section {\bf 3}, we consider the Husimi lattice with $q=4$ and find
recursion relations for numbers of allowed configurations in the SOC state.
In section {\bf 4}, we compute exactly the distribution of height 
probabilities in
this particular case.  In section  {\bf 5}, we generalize our results to an
arbitrary coordination number of the lattice.

\section{Lattice and Model.}

As a basic building block for constructing a {\sl cactus} or a
{\sl pure Husimi tree}~\cite{EF}, we will take $\gamma$ triangle
plaquettes joined at the {\sl base site} (root), see Fig.1a. This
basic building block  will be called the {\sl first-\-generation
branch}. To the {\sl second-\-generation branch}, we attach $2\gamma$
basic building blocks at each free site of the first-\-generation
branch (Fig.1b).
Continuing this process we develop higher-generation branches.
Then, at the final step, we take three $n$th-\-generation
branches and connect their base sites by the triangle plaquette.
As a result, we will get the cactus with the coordination number
$q=2(\gamma+1)$.

Let as define  ASM on this connected graph of $N$ sites
as follows: to each site $i$ $(1\leq i\leq N)$ we associate
an integer $z_i$ $(1\leq z_i\leq q)$ which is the height
of a column of sand grains. The evolution of the system is
specified by two rules:

\bigskip
{\bf i.} Addition of a sand grain at a randomly chosen site $i$ 
increases $z_i$ by 1.

{\bf ii.} The site $i$ topples if the height $z_i$ exceeds a 
critical value
$z_c=q$ and sand grains drop on the nearest neighbors.

\bigskip
The number of surface sites of the Husimi tree is comparable
with the interior ones. Hence, the calculation of the thermodynamic
limit of the bulk properties
requires special care. In our work we define the height distribution
of sand grains for the sites deep inside the tree. Using these
interior sites one can construct an infinite lattice,
for they have the same features. Therefore, we will consider
the problem on the Husimi lattice rather than on the Husimi tree.

Any configuration $\{z_i\}$ on the Husimi tree in which 
$1\leq z_i\leq q$
is a stable configuration under the toppling rule. These configurations
can be divided into two classes: allowed and forbidden
configurations~\cite{Dhar}.

In the SOC state, only allowed configurations have a nonzero
 probability.
Any subconfiguration of heights $F$ on a finite connected set
of sites is forbidden if
$$
z_i\leq q_i,
$$
where $q_i$ is a coordination number of a site $i$ in the given
subconfiguration $F$~\cite{Dhar}.

In turn, we can divide the allowed
configurations on an $n$th-generation branch of the Husimi tree
into three nonoverlapping classes:
{\sl weakly allowed of the type 1 ($W_1$),
weakly allowed of the type 2 ($W_2$) and
strongly allowed ($S$)} configurations.

Consider an allowed subconfiguration $C$
on the $n$th-generation branch $G_n$ with a root
$a$ (Fig.2a). The coordination number of the root $a$ is $q-2$. 
Adding
a vertex $b$ to the $G_n$, one defines a
subgraph $G'=G_n \cup b$. If the subconfiguration $C'=C \cup b$ 
with $z_b=1$
on the $G'$ is forbidden, then
$C$ is called the weakly allowed subconfiguration of  type {1} 
($W_1$). Thus, $W_1$ can be locked by one bond.

Now add two vertices $b$ and $d$ to the $G_n$ and
consider a subconfiguration $C''=C \cup b \cup d$
on  $G''=G_n \cup b  \cup d$ (Fig.2b).
If the subconfiguration $C''=C \cup b \cup d$ with $z_b=1, z_d=1$
on the $G''$ is forbidden, then
$C$ is called the weakly allowed subconfiguration of type {2} ($W_2$).

Any allowed subconfigurations defined on the $n$th-generation 
branches
that cannot be locked by one bond or by two bonds form
a strongly allowed class ($S$).

It is important to note that any subconfiguration of the $W_1$
type is also of the $W_2$ type. To obtain the nonoverlapping
classes, we always check the subconfiguration first to
belong to the $W_1$ type and only after that to the $W_2$ type.

We will start with mentioning some forbidden subconfigurations
that can occur on the Husimi tree.

Let us first consider the
subconfiguration shown in  Fig.3a. It is easy to check that
$C=C_1 \cup C_2$ will be forbidden if both $C_1$ and $C_2$
are of the $W_1$ type. The next example of the forbidden 
subconfiguration
is shown in Fig.3b where configurations  $C_1$, $C_2$ and $C_3$
should be of the $W_1$ or $W_2$ type.

\newpage
\section{Recursion relations.}

\setcounter{equation}0

In this section, for simplicity, we will consider a Husimi
tree with $\gamma=1$ $(q=4)$. The $n$th-generation branch $G_n$
with a root $a$ (Fig.4) consists of two $(n-1)$th-generation branches
$G_{n-1}^{(1)}$ and $G_{n-1}^{(2)}$ with roots $a_1$ and $a_2$, 
respectively.
Vertices $a_1$ and $a_2$ are nearest
neighbors of the root $a$. Let $N_{W_1}(G_n,i)$, $N_{W_2}(G_n,i)$
and $N_{S}(G_n,i)$ be the numbers of distinct $W_1$, $W_2$ and $S$
type subconfigurations on the $G_n$ with a given height $z_a=i$.
Let us also introduce
\begin{equation}
\label{f3.1}
N_{W_1}(G_n)=\sum\limits_{i=1}^4N_{W_1}(G_n,i),
\end{equation}

\begin{equation}
\label{f3.2}
N_{W_2}(G_n)=\sum\limits_{i=1}^4N_{W_2}(G_n,i),
\end{equation}

\begin{equation}
\label{f3.3}
N_S(G_n)=\sum\limits_{i=1}^4N_S(G_n,i).
\end{equation}

The number of allowed subconfigurations on the $G_n$
with a given root is
\begin{equation}
\label{f3.4}
N(G_n,i)=N_{W_1}(G_n,i)+N_{W_2}(G_n,i)+N_S(G_n,i).
\end{equation}
At the same time, $N_{W_1}(G_n)$, $N_{W_2}(G_n)$ and $N_S(G_n)$
can be expressed in terms
of the full numbers of allowed subconfigurations
 on the two $(n-1)$th-generation branches
$G^{(1)}_{n-1}$ and $G^{(2)}_{n-1}$:
$$
N_{W_1}(G_n)=N_S(G_{n-1}^{(1)})N_S(G_{n-1}^{(2)})+N_S(G_{n-1}^{(1)})
N_{W_1}(G_{n-1}^{(2)})+N_{W_1}(G_{n-1}^{(1)})N_S(G_{n-1}^{(2)})+
$$
\begin{equation}
\label{f3.6}
N_S(G_{n-1}^{(1)})N_{W_2}(G_{n-1}^{(2)})+N_{W_2}(G_{n-1}^{(1)})
N_S(G_{n-1}^{(2)})+N_{W_1}(G_{n-1}^{(1)})N_{W_2}(G_{n-1}^{(2)})+
\end{equation}
$$
N_{W_2}(G_{n-1}^{(1)})N_{W_1}(G_{n-1}^{(2)})+N_{W_2}(G_{n-1}^{(1)})
N_{W_2}(G_{n-1}^{(2)}),
$$

$$
N_{W_2}(G_n)=N_S(G_{n-1}^{(1)})N_S(G_{n-1}^{(2)})+N_S(G_{n-1}^{(1)})
N_{W_1}(G_{n-1}^{(2)})+N_{W_1}(G_{n-1}^{(1)})N_S(G_{n-1}^{(2)})+
$$
\begin{equation}
\label{f3.7}
N_S(G_{n-1}^{(1)})N_{W_2}(G_{n-1}^{(2)})+N_{W_2}(G_{n-1}^{(1)})
N_S(G_{n-1}^{(2)})+N_{W_1}(G_{n-1}^{(1)})N_{W_2}(G_{n-1}^{(2)})+
\end{equation}
$$
N_{W_2}(G_{n-1}^{(1)})N_{W_1}(G_{n-1}^{(2)})+N_{W_2}(G_{n-1}^{(1)})
N_{W_2}(G_{n-1}^{(2)}),
$$

$$
N_S(G_n)=2N_S(G_{n-1}^{(1)})N_S(G_{n-1}^{(2)})+N_S(G_{n-1}^{(1)})
N_{W_1}(G_{n-1}^{(2)})+N_{W_1}(G_{n-1}^{(1)})N_S(G_{n-1}^{(2)})+
$$
\begin{equation}
\label{f3.8}
2N_S(G_{n-1}^{(1)})N_{W_2}(G_{n-1}^{(2)})+2N_{W_2}(G_{n-1}^{(1)})
N_S(G_{n-1}^{(2)}).
\end{equation}

\bigskip
As we can see, expressions~(\ref{f3.6}) and~(\ref{f3.7}) are equal
to each other.
We will prove in section {\bf 5} that this holds for all $q$

Let us define

\begin{equation}
\label{f3.9}
X=\frac{N_W}{N_S}\,,
\end{equation}
where $N_W\equiv N_{W_1}=N_{W_2}$.

From~(\ref{f3.6})-(\ref{f3.8}) one may find the following
recursion relation:
\begin{equation}
\label{f3.10}
X(G_n)=\frac{%
1+2X(G_{n-1}^{(1)})+2X(G_{n-1}^{(2)})+3X(G_{n-1}^{(1)})
X(G_{n-1}^{(2)})}{2+3X(G_{n-1}^{(1)})+3X(G_{n-1}^{(2)})}\,.
\end{equation}

If we consider graphs $G^{(1)}_{n-1}$ and $G^{(2)}_{n-1}$
to be isomorphic, then $N(G^{(1)}_{n-1})=N(G^{(2)}_{n-1})$
and equation~(\ref{f3.10}) simplifies to
\begin{equation}
\label{f3.11}
X(G_n)=\frac{1}{2}\left(1+X(G_{n-1})\right)\,.
\end{equation}

With the initial condition $X(G_0)=1/2$, equation~(\ref{f3.11})
has a simple solution
\begin{equation}
\label{f3.12}
X(G_n)=1-2^{-(n+1)}\,.
\end{equation}

On the infinite generation branches $(n\rightarrow \infty)$
the ratio of the number of the $W_1$ type configurations or
of the $W_2$ type to the strongly allowed ones tends to 1.

\section{Distribution of height probabilities.}

\setcounter{equation}0

Consider now a randomly chosen site $O$ deep inside
the Husimi tree (Fig.5). The nearest neighbors of the
site $O$ are  roots of the four $n$th-generation branches.
For a given value $z_0=i$ $(1 \leq i \leq 4)$ at the site $O$,
the number of allowed configurations $N(i)$ is expressed via
$N(G_n^{(1)})$, $N(G_n^{(2)})$, $N(G_n^{(3)})$, $N(G_n^{(4)})$,
where again 
$N(G_n^{(\alpha )})=\sum\nolimits_{i=1}^4N(G_n^{(\alpha )},i)$,
$\alpha =1,2,3,4$.

If $i=1$, then each allowed configuration on the branches
$G^{(1)}_n$, $G^{(2)}_n$, $G^{(3)}_n$, $G^{(4)}_n$ cannot be of 
the $W_1$ type.
It is also evident that two $W_2$ type configurations
cannot occur on the neighboring branches $G^{(1)}_n$ and $G^{(2)}_n$
or $G^{(3)}_n$ and $G^{(4)}_n$. Hence, the number $N(1)$ of allowed
configurations with isomorphic branches is
\begin{equation}
\label{f4.2}
N(1)=[1+4X+4X^2]\prod\limits_{\alpha=1}^{4}N_S(G^{(\alpha)})\,.
\end{equation}

Arguing similarly, one can get the following expressions for the 
numbers
of allowed configurations for a given value at the site $i$
\begin{equation}
\label{f4.3}
N(2)=[1+8X+12X^2]\prod\limits_{\alpha=1}^{4}N_S(G^{(\alpha)})\,,
\end{equation}

\begin{equation}
\label{f4.4}
N(3)=[1+8X+22X^2+12X^3]\prod\limits_{\alpha=1}^{4}N_S(G^{(\alpha)})\,,
\end{equation}

\begin{equation}
\label{f4.5}
N(4)=[1+8X+22X^2+24X^3]\prod\limits_{\alpha=1}^{4}N_S(G^{(\alpha)})\,.
\end{equation}

The probability $P(i)$ of having the height $i$ at site $O$ is
\begin{equation}
\label{f4.6}
P(i)=\frac{N(i)}{N_{\mbox{\scriptsize total}}}\,,
\end{equation}
where $N_{\mbox{\scriptsize total}}=\sum_{i=1}^4N(i)$
is the total number of allowed configurations on the
Husimi tree.

The system can reach the SOC state only in the thermodynamic limit.
For the sites far from the surfaces
in this limit $(n\rightarrow \infty)$ we have $X=1$. Thus,
from~(\ref{f4.2})-(\ref{f4.6}) we get
\begin{equation}
\label{f4.7}
P(1)=\frac
9{128}\,,\quad P(2)=\frac{21}{128}\,,\quad P(3)=\frac{43}{128}\,,
\quad P(4)=\frac{55}{128}\,.
\end{equation}

The obtained values~(\ref{f4.7}) for the height probabilities
characterize the SOC state of the ASM on the Husimi lattice with
$q=4$.

\bigskip
\section{Generalization to the Husimi lattice with an
arbitrary coordination number $q$.}

\setcounter{equation}0

We now intend to generalize results obtained in
two previous sections. Before the recursion relations
and expressions for the height probabilities will be
written out, we want to prove that the number of the
$W_1$ type configurations and the number of the $W_2$ type ones
is equal on the $n$th-generation branch of the Husimi tree.

The numbers of weakly allowed configurations of
both the types can be written as
\begin{eqnarray}
\label{f5.1}
N_{W_1}(G_n) = & N_{W_1}(G_n,1)+& N_{W_1}(G_n,2)+\cdots +
N_{W_1}(G_n,q),  \\
N_{W_2}(G_n) = &                & N_{W_2}(G_n,2)+\cdots +
N_{W_2}(G_n,q-1) + N_{W_2}(G_n,q).
\label{f5.2}
\end{eqnarray}

It is easy to see that $N_{W_1}(G_n,1)=N_{W_2}(G_n,q)=0$.
By definition, the $W_1$ type configurations
can be locked by one bond.
When the height of the root is increased by 1, it becomes of
$W_2$ type. Similarly, decreasing the height of the root
of $W_2$ type configuration by 1, we get the $W_1$ type.
Thus for each configuration of the $W_1$ type
there is a unique configuration of the $W_2$ type
and vice versa. Therefore, we can conclude that~(\ref{f5.1})
and~(\ref{f5.2}) are equal.

For an arbitrary coordination number of the Husimi lattice,
equation~(\ref{f3.11}) generalizes to
\begin{equation}
\label{f5.3}
X(G_n)=\frac{1}{2\gamma}\left(1+X(G_{n-1})\right).
\end{equation}
and with the initial condition $X(G_0)=1/(2\gamma)$, it has the solution
\begin{equation}
\label{f5.4}
X(G_n)=\frac 1{2\gamma -1}-\frac{1}{2\gamma -1}(2\gamma )^{-(n+1)}\,.
\end{equation}

For large enough $n$ we have
$X(G_n)\stackrel{n\to\infty}{=}1/{(2\gamma-1)}$.

In the same way as for the $q=4$ case, we can calculate the
height probabilities of the sites deep inside the
lattice
\begin{eqnarray}
\label{f5.5}
P(i) & = & \frac{(2\gamma +1)^{\gamma +1}}{2^{\gamma +2}(\gamma +1)^2
(\gamma+3)^\gamma }\times \nonumber       \\
     &   & \sum\limits_{n_1=0}^{i-1}
\sum\limits_{n_2=0}^{(i-1)/2~i\mbox{\scriptsize --odd} \atop
                     (i-2)/2~i\mbox{\scriptsize --even}}
{\frac{(\gamma +1)!}
{(\gamma +1-n_1-n_2)!\cdot n_1!\cdot n_2!}}
2^{n_1}3^{n_2}(2\gamma+1)^{-(n_1+n_2)}\,,
\end{eqnarray}
where $n_1+n_2\leq \gamma+1$ and
$1 \leq i \leq q$.

The probability of the height to be equal 1 has a simple form
\begin{equation}
\label{f5.6}
P(1)=\frac{(2\gamma +1)^{\gamma +1}}{2^{\gamma +2}(\gamma +1)^2(\gamma
+3)^\gamma }\,,
\end{equation}
and tends to 0 when $\gamma\rightarrow \infty$.

In this article, we have exactly calculated  the distribution of height
probabilities in the Self-Organized Critical state of the ASM on
the Husimi lattice of triangles. The next step of our investigations
will be the Husimi lattice with square plaquettes where we intend to
derive exact expressions for two-point correlation functions
and critical exponents of avalanches. The choice of the Husimi lattice
gives us hope to find the relationship between the chaos and SOC state,
as some spin models formulated on this lattice show the chaotic
behavior~\cite{Mon,Ners}.

\section*{Acknowledgments}

We are grateful to N.S.~Ananikian and V.B.~Priezzhev
for fruitful discussions.

One of us (R.R.S.) was partially supported by  German
Bundesministerium f\"ur Fors\-chung and Technologie under the
grant N 211-5291 YPI.

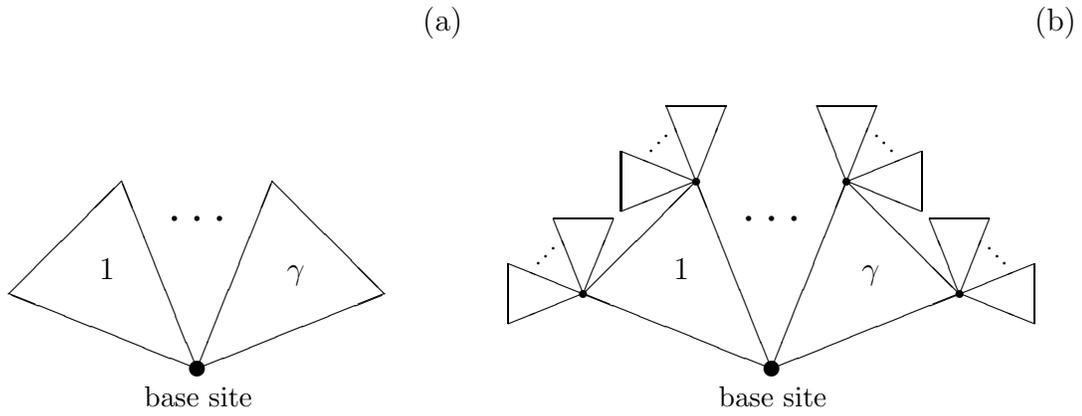
\begin{figure}[p]
\unitlength 1mm
\begin{picture}(75,60)
\put(65,55){\mbox{(a)}}
%
%
%
\put(35,10){\line(-5,2){25}}
\put(35,10){\line(-2,5){10}}
\put(10,20){\line(1,1){15}}
\put(22,22){$1$}
%
%
\put(35,10){\line(2,5){10}}
\put(35,10){\line(5,2){25}}
\put(45,35){\line(1,-1){15}}
\put(47,22){$\gamma$}
\multiput(32,30)(3,0){3}{\circle*{0.8}}
\put(35,10){\circle*{2}}
\put(28,5){\mbox{\small base site}}
%
\end{picture}
%
\begin{picture}(75,75)
\put(70,55){\mbox{(b)}}
%
%
%
\put(35,10){\line(-5,2){25}}
\put(35,10){\line(-2,5){10}}
\put(10,20){\line(1,1){15}}
\put(22,22){$1$}
\put(10,20){\line(-5,-2){10}}
\put(10,20){\line(-5,2){10}}
\put(0,16){\line(0,1){8}}
\multiput(4,24)(1,1){3}{\circle*{0.5}}
\put(10,20){\circle*{1}}
%
\put(10,20){\line(-2,5){4}}
\put(10,20){\line(2,5){4}}
\put(6,30){\line(1,0){8}}
\put(25,35){\line(-5,-2){10}}
\put(25,35){\line(-5,2){10}}
\put(15,31){\line(0,1){8}}
\multiput(19,39)(1,1){3}{\circle*{0.5}}
\put(25,35){\circle*{1}}
%
\put(25,35){\line(-2,5){4}}
\put(25,35){\line(2,5){4}}
\put(21,45){\line(1,0){8}}
%
%
\put(35,10){\line(2,5){10}}
\put(35,10){\line(5,2){25}}
\put(45,35){\line(1,-1){15}}
\put(47,22){$\gamma$}
\put(45,35){\line(-2,5){4}}
\put(45,35){\line(2,5){4}}
\put(41,45){\line(1,0){8}}
\multiput(49,41)(1,-1){3}{\circle*{0.5}}
\put(45,35){\circle*{1}}
%
%
\put(45,35){\line(5,2){10}}
\put(45,35){\line(5,-2){10}}
\put(55,39){\line(0,-1){8}}
\put(60,20){\line(-2,5){4}}
\put(60,20){\line(2,5){4}}
\put(56,30){\line(1,0){8}}
\multiput(64,26)(1,-1){3}{\circle*{0.5}}
\put(60,20){\circle*{1}}
%
%
\put(60,20){\line(5,2){10}}
\put(60,20){\line(5,-2){10}}
\put(70,24){\line(0,-1){8}}
\multiput(32,30)(3,0){3}{\circle*{0.8}}
\put(35,10){\circle*{2}}
\put(28,5){\mbox{\small base site}}
%
\end{picture}
\caption{(a) A first-generation branch which consists of $\gamma$
triangle plaquettes joined at the {\sl base site}.
(b) A second-generation branch.}
\end{figure}

\begin{figure}[p]
\unitlength 1mm
\begin{picture}(75,75)
\put(65,70){\mbox{(a)}}
%
%
%
\put(35,25){\line(-5,2){25}}
\put(35,25){\line(-2,5){10}}
\put(10,35){\line(1,1){15}}
\put(22,37){$1$}
%
%
\put(0,35){\oval(20,10)[r]}
%
%
\put(25,60){\oval(10,20)[b]}
%
%
\put(35,25){\line(2,5){10}}
\put(35,25){\line(5,2){25}}
\put(45,50){\line(1,-1){15}}
\put(47,37){$\gamma$}
%
%
\put(45,60){\oval(10,20)[b]}
%
%
\put(70,35){\oval(20,10)[l]}
\multiput(32,45)(3,0){3}{\circle*{0.8}}
\put(35,25){\circle*{2}}
\put(30,22){$a$}
\multiput(35,25)(0,-6){4}{\line(0,-1){3}}
\put(35,4){\circle*{2}}
\put(30,2){$b$}
%
\end{picture}
%
\begin{picture}(75,75)
\put(65,70){\mbox{(b)}}
%
%
%
\put(35,25){\line(-5,2){25}}
\put(35,25){\line(-2,5){10}}
\put(10,35){\line(1,1){15}}
\put(22,37){$1$}
%
%
\put(0,35){\oval(20,10)[r]}
%
%
\put(25,60){\oval(10,20)[b]}
%
%
\put(35,25){\line(2,5){10}}
\put(35,25){\line(5,2){25}}
\put(45,50){\line(1,-1){15}}
\put(47,37){$\gamma$}
%
%
\put(45,60){\oval(10,20)[b]}
%
%
\put(70,35){\oval(20,10)[l]}
\multiput(32,45)(3,0){3}{\circle*{0.8}}
\put(35,25){\circle*{2}}
\put(30,22){$a$}
\multiput(35,25)(-2.228,-5.571){4}{\line(-2,-5){1.5}}
\put(26.5,4){\circle*{2}}
\put(22,2){$b$}
\multiput(35,25)(2.228,-5.571){4}{\line(2,-5){1.5}}
\put(43.5,4){\circle*{2}}
\put(46,2){$d$}
%
\end{picture}
\caption{(a) A $n$th-generation branch $G_n$ and vertex $b$ form a
subgraph $G'$. (b) Now two vertexes $b$ and $d$ and the $G_n$
form a subgraph $G''$.}
\end{figure}
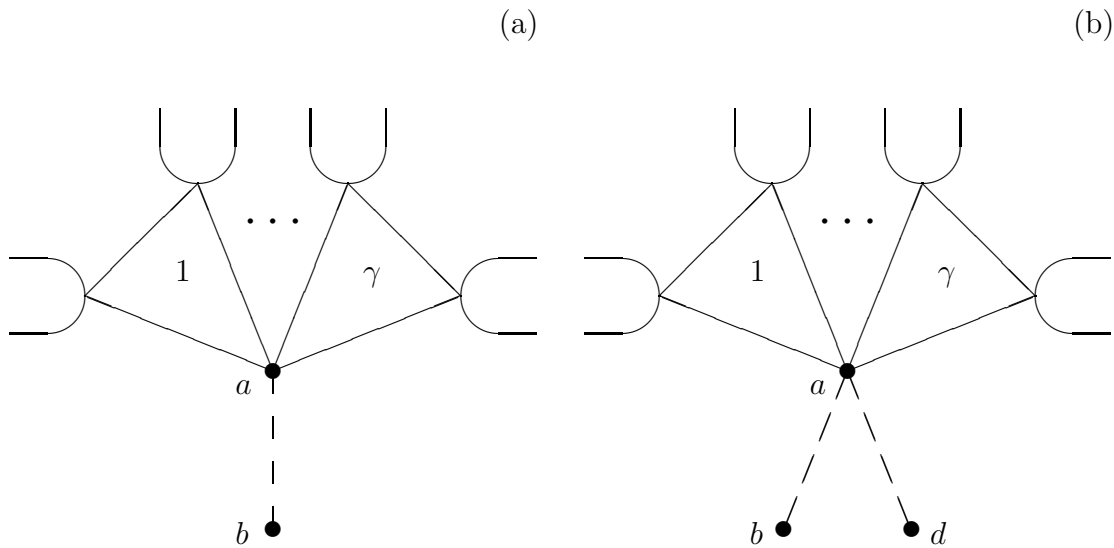

\begin{figure}[p]
\unitlength 1mm
\begin{picture}(75,60)
\put(65,55){\mbox{(a)}}
\put(25,30){\line(1,0){20}}
%
%
\put(15,30){\oval(20,10)[r]}
\put(15,29){$C_1$}
%
%
\put(55,30){\oval(20,10)[l]}
\put(51,29){$C_2$}
\put(25,30){\circle*{2}}
\put(27,32){$a$}
\put(45,30){\circle*{2}}
\put(42,32){$b$}
%
\end{picture}
\begin{picture}(75,60)
\put(65,55){\mbox{(b)}}
%
%
\put(25,40){\line(1,0){20}}
\put(25,40){\line(1,-2){10}}
\put(45,40){\line(-1,-2){10}}
%
%
\put(14.5,50){\oval(25,25)[rb]}
\put(14.5,46){$C_1$}
%
%
\put(55.5,50){\oval(25,25)[lb]}
\put(51,46){$C_2$}
%
%
\put(35,10){\oval(10,20)[t]}
\put(33,10){$C_3$}
\put(25,40){\circle*{2}}
\put(29,42){$a$}
\put(45,40){\circle*{2}}
\put(39,42){$b$}
\put(35,20){\circle*{2}}
\put(34,24.5){$d$}
%
\end{picture}
\caption{Two examples of forbidden subconfigurations that can occur
on the Husimi tree.}
\end{figure}
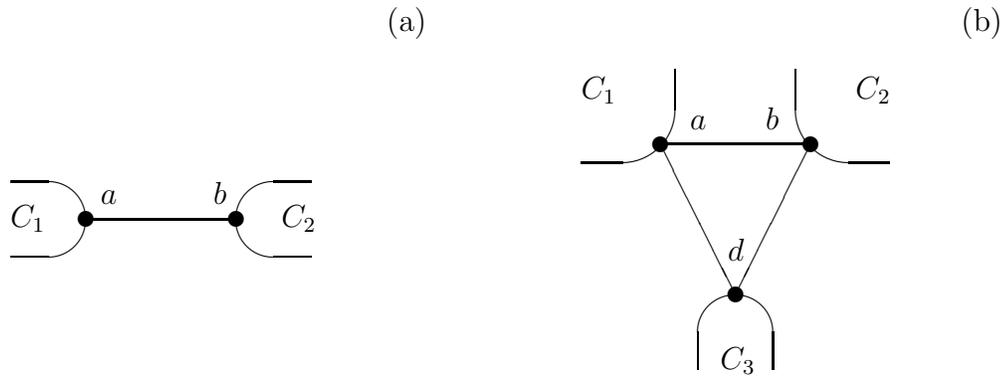

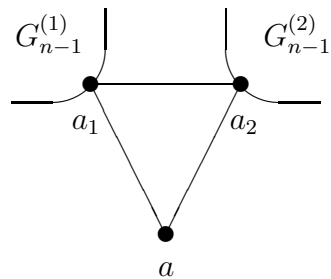
\begin{figure}[p]
\begin{center}
\unitlength 1mm
\begin{picture}(75,60)
%
%
%
\put(25,35){\line(1,0){20}}
\put(25,35){\line(1,-2){10}}
\put(45,35){\line(-1,-2){10}}
%
%
\put(14.5,45){\oval(25,25)[rb]}
\put(15,40){$G_{n-1}^{(1)}$}
%
%
\put(55.5,45){\oval(25,25)[lb]}
\put(48,40){$G_{n-1}^{(2)}$}
\put(25,35){\circle*{2}}
\put(22.5,29){$a_1$}
\put(45,35){\circle*{2}}
\put(43.5,29){$a_2$}
\put(35,15){\circle*{2}}
\put(34,9.5){$a$}
%
\end{picture}

\end{center}
\caption{The $n$th-generation branch $G_n$ with
two nearest $(n-1)$th-generation branches $G_{n-1}^{(1)}$ and
$G_{n-1}^{(2)}$.}
\end{figure}

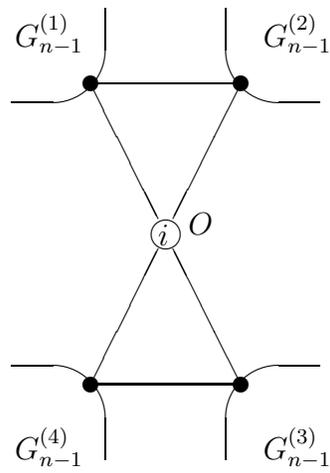
\begin{figure}[p]
\begin{center}
\unitlength 1mm
\begin{picture}(75,75)
%
%
%
\put(25,60){\line(1,0){20}}
\put(25,60){\line(1,-2){9}}
\put(45,60){\line(-1,-2){9}}
\put(25,60){\circle*{2}}
\put(45,60){\circle*{2}}
%
%
\put(14.5,70){\oval(25,25)[rb]}
\put(15,65){$G_{n-1}^{(1)}$}
%
%
\put(55.5,70){\oval(25,25)[lb]}
\put(48,65){$G_{n-1}^{(2)}$}
\put(35,40){\circle{4}}
\put(38,40){$O$}
\put(34,38.5){$i$}
%
%
%
\put(25,20){\line(1,0){20}}
\put(25,20){\line(1,2){9}}
\put(45,20){\line(-1,2){9}}
\put(25,20){\circle*{2}}
\put(45,20){\circle*{2}}
%
%
\put(14.5,10){\oval(25,25)[rt]}
\put(15,10){$G_{n-1}^{(4)}$}
%
%
\put(55.5,10){\oval(25,25)[lt]}
\put(48,10){$G_{n-1}^{(3)}$}
%
\end{picture}

\end{center}
\caption{A site $O$ with height $i$ is located deep inside the 
lattice
and surrounded by the four $n$th-generation branches
$G_n^{(1)}, G_n^{(2)}, G_n^{(3)}, G_n^{(4)}$.}
\end{figure}

\end{document}